\documentclass[aps, twocolumn, prl, amssymb, amsmath, superscriptaddress,preprintnumbers]{revtex4-1}
\UseRawInputEncoding
\usepackage{graphicx}
\usepackage{dcolumn}
\usepackage{bm}
\usepackage{amsmath}
\usepackage{subfigure}
\usepackage{amsfonts}
\usepackage{xcolor}
\usepackage{appendix}
\usepackage{multirow}
\usepackage{booktabs}
\usepackage{tabularx}
\usepackage[colorlinks,
            linkcolor=blue,
            anchorcolor=red,
            citecolor=blue
            ]{hyperref}
\usepackage{amsthm}
\usepackage{setspace}
\usepackage{lineno}

\begin{document}
%\begin{CJK*}{UTF8}{gbsn}
%\linenumbers

\title{Sequential generalized measurements: Asymptotics, typicality and emergent projective measurements}

\author{Wen-Long Ma}
\email{wenlongma@semi.ac.cn}
\affiliation{State Key Laboratory of Superlattices and Microstructures, Institute of Semiconductors, Chinese Academy of Sciences, Beijing 100083, China}
\affiliation{Center of Materials Science and Opto-Electronic Technology, University of Chinese Academy of Sciences, Beijing 100049, China}
%\affiliation{Hefei National Laboratory, Hefei 230088, China}
\author{Shu-Shen Li}
\affiliation{State Key Laboratory of Superlattices and Microstructures, Institute of Semiconductors, Chinese Academy of Sciences, Beijing 100083, China}
\affiliation{Center of Materials Science and Opto-Electronic Technology, University of Chinese Academy of Sciences, Beijing 100049, China}
\author{Ren-Bao Liu}
%\email{rbliu@cuhk.edu.hk}
\affiliation{Department of Physics, Centre for Quantum Coherence, and The Hong Kong Institute of Quantum Information Science and Technology, The Chinese University of Hong Kong, Shatin, New Territories, Hong Kong, China}

\date{\today }

\begin{abstract}
The relation between projective measurements and generalized quantum measurements is a fundamental problem in quantum physics, and clarifying this issue is also important to quantum technologies. While it has been intuitively known that projective measurements can be constructed from sequential generalized or weak measurements, there is still lack of a proof of this hypothesis in general cases. Here we prove it from the perspective of quantum channels. We show that projective measurements naturally arise from sequential generalized measurements in the asymptotic limit. Specifically, a selective projective measurement arises from a set of typical sequences of selective generalized measurements. We provide an explicit scheme to construct projective measurements of a quantum system with sequential generalized measurements. Remarkably, a single ancilla qubit is sufficient to mediate sequential generalized measurements for constructing arbitrary projective measurements of a generic system. %As an example, we present a protocol to measure the modular excitation number of a bosonic mode with an ancilla qubit.

\end{abstract}

\maketitle

Quantum measurements retrieve classical information from quantum states \cite{Wiseman2010,Jacobs2014}, and are particularly important to quantum technologies \cite{Nielsen2010}. The traditional description of measurement in quantum mechanics is through projective measurements (PMs) of observables represented by Hermitian operators \cite{Peres2006}. Measuring an observable corresponds to statistically projecting the quantum state to one of the orthogonal eigenspaces of this observable. PMs appear most commonly in quantum foundation and quantum information theory, and are widely useful for initialization and readout of quantum systems in quantum technologies \cite{Eizerman2004,Vamivakas2010,Neumann2010,Morello2010,Jiang2009,Nakajima2017,West2019}.

There exist more general quantum measurements, called generalized measurements (GMs) described by positive-operator-valued measures (POVMs) \cite{Kraus1983,Andersson2008,Chen2018,Chen2019,Cheong2012}. GMs can outperform PMs in many tasks in quantum technologies, such as quantum tomography \cite{Renes2004} and quantum state discrimination or estimation \cite{Bergou2010,Derka2001}. Moreover, continuous or sequential GMs can be exploited for monitoring and maneuvering quantum evolutions \cite{Jacobs2006,Gurvitz1997,Ashhab2009,Korotkov2001,Blok2006,Jordan2006,Chantasri2013,Chantasri2015,Presilla1996,Diosi2016}. In particular, weak measurements can extract partial information without projections, and therefore can help realize optimal qubit tomography \cite{Shojaee2018}, reconcile measurement incompatibility \cite{Monroe2021,Guhne2022} and extract arbitrary bath correlations \cite{Wang2019,Pfender2019,Wu2022}.

Substantial efforts have been devoted to illustrating the relation between PMs and GMs. A celebrated result is Naimark's theorem \cite{Peres2006}, implying that any GM can be implemented as a PM on an enlarged Hilbert space. The measurement statistics of GMs can also be simulated by PMs with classical randomness or postselection \cite{Oszmaniec2017,Oszmaniec2019,Singal2022}. In the opposite direction, it has been argued that sequential GMs can generate PMs by analysing the gradual state collapse \cite{Brun2002,Lidar2013,Oreshkov2005,Varbanov2007}, the statistics of measurement results \cite{Ma2018,Rao2019,Liu2017} and saturation of knowledge \cite{Haapasalo2016}. However, to our knowledge, the general relation between PMs and sequential GMs still remains elusive.

In this paper, we prove that PMs can emerge from sequential GMs in the asymptotic limit, when the measurement operators are normal and commuting with each other. The proof is based on the observation that projections are fixed points of the quantum channels for such GMs. Moreover, from the theory of classical typicality, we find that different selective PMs arise from different sets of typical sequences of selective GMs. These results completely characterize the structures of sequential GMs with normal and commuting measurement operators. We further present a general scheme to realize such GMs with a single qubit ancilla, and show that sequential GMs can simulate arbitrary PMs for arbitrary finite-dimensional quantum systems. The scheme will be useful for initialization, readout and feedback control of a quantum system. As an example, we provide a protocol to measure the modular excitation numbers of an infinite-dimensional bosonic mode with an ancilla qubit, which are the error syndromes of several bosonic quantum error correction codes.

\begin{figure*}
\includegraphics[width=6.0in]{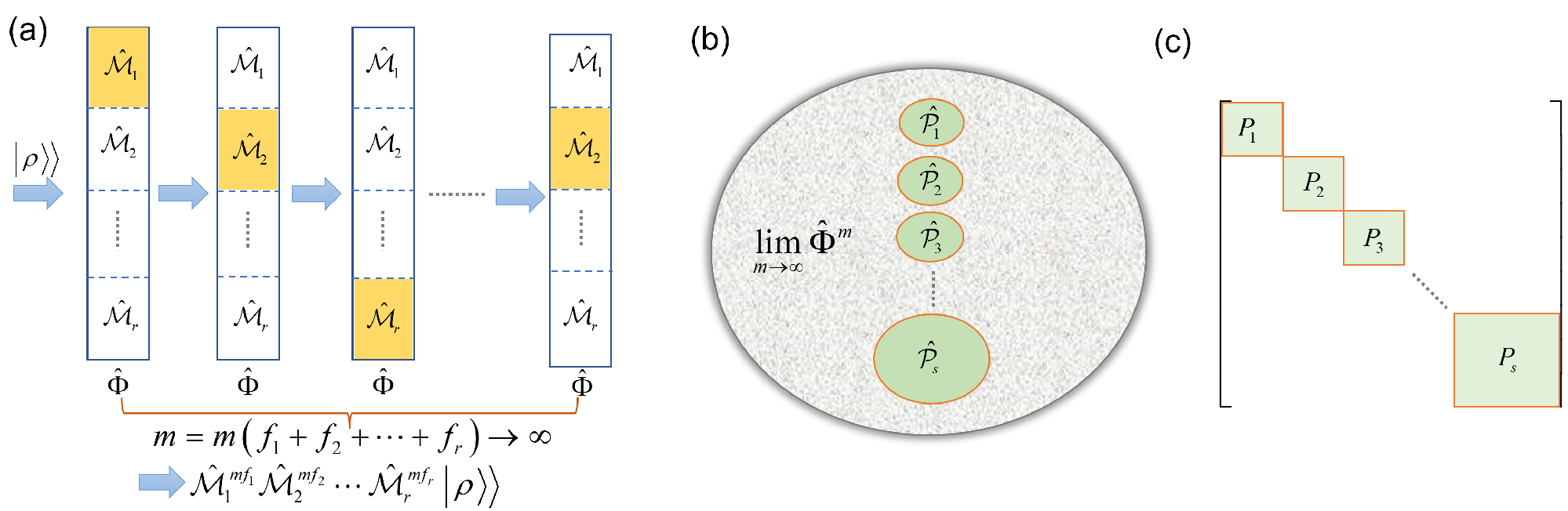}
\caption{(a) Schematic of sequential non-selective GMs and sequences of selective GMs in the asymptotic limit. (b) Emergent PMs arising from summation over the sets of typical sequences of selective GMs. (c) The emergent projections in the operator space of the quantum system. }
\label{SGM}
\end{figure*}

\textit{GMs and quantum channels}. For a $d$-level quantum system, a $r$-outcome POVM is a set of positive semidefinite operators acting in the Hilbert space that sum to the identity, $\sum_{\alpha=1}^r M_{\alpha}^{\dagger}M_{\alpha}=\mathbb{I}$. The $\alpha$th outcome is obtained with probability Tr$(M_{\alpha}^{\dagger}M_{\alpha}\rho)$ with $\rho$ being the density matrix. A GM is characterized by a POVM and the set of measurement operators $\{M_{\alpha}\}_{\alpha=1}^r$. The state change induced by a GM is described by a completely positive and trace-preserving (CPTP) map or a quantum channel \cite{Kraus1983,Caruso2014},
\begin{equation}\label{}
    \Phi(\rho)=\sum_{\alpha=1}^{r}\mathcal{M}_{\alpha}\rho=\sum_{\alpha=1}^{r}M_{\alpha}\rho M_{\alpha}^{\dagger},
\end{equation}
where $\mathcal{M}_{\alpha}=M_{\alpha}(\cdot)M_{\alpha}^{\dagger}$ is a superoperator acting in the operator space of the quantum system, representing a trace-nonincreasing and completely positive (CP) map corresponding to the $\alpha$th outcome. The set of superoperators $\{\mathcal{M}_{\alpha}\}_{\alpha=1}^r$ form a quantum instrument \cite{Davies1970,Ozawa1984}, which belongs to a class of quantum channels that can include both classical and quantum outputs. Hereafter we define a non-selective GM as the channel $\Phi=\sum_{\alpha=1}^r \mathcal{M}_{\alpha}$, and a selective GM as a specific CP map $\mathcal{M}_{\alpha}$. 

Quantum channels have natural matrix representations in the Hilbert-Schmidt (HS) space of the quantum system \cite{Bengtsson2017,SI}. While the density matrices are operators in the Hilbert space with an orthonormal basis $\{|a\rangle\}_{a=1}^d$, they are turned into vectors in the HS space, i.e., $\rho=\sum_{a,b=1}^{d}\rho_{ab}|a\rangle\langle b|\leftrightarrow |\rho\rangle\rangle=\sum_{a,b=1}^{d}\rho_{ab}|ab\rangle\rangle$, such that $X\rho Y\leftrightarrow X\otimes Y^{T}|\rho\rangle\rangle$ with $X$, $Y$ being operators acting in the Hilbert space and $Y^{T}$ being the transpose of $Y$. The inner product in the HS space is defined as $\langle\langle\sigma|\rho\rangle\rangle={\rm Tr} [\sigma^{\dagger}\rho]$. The quantum channel is a linear operator acting in the HS space,
\begin{equation}\label{}
    \hat{\Phi}|\rho\rangle\rangle=\sum_{\alpha=1}^{r}\hat{\mathcal{M}}_{\alpha}|\rho\rangle\rangle=\sum_{\alpha=1}^{r}M_{\alpha}\otimes M_{\alpha}^{*}|\rho\rangle\rangle,
\end{equation}
where $M_{\alpha}^{*}$ is the complex conjugate of $M_{\alpha}$. Note that we add hats for operators acting in the HS space, to distinguish them from the corresponding superoperators acting in the operator space of the quantum system. With the HS space, the probability to get the $\alpha$th outcome is $\langle\langle \mathbb{I}|\hat{\mathcal{M}}_{\alpha}|\rho\rangle\rangle={\rm Tr}(M_{\alpha}\rho M_{\alpha}^{\dagger})$.

\textit{GMs with normal and commuting measurement operators}. We assume that the set of measurement operators $\{M_{\alpha}\}_{\alpha=1}^r$ are normal and commuting with each other, i.e., $[M_{\alpha}, M_{\alpha}^{\dagger}]=[M_{\alpha}, M_{\beta}]=0$ for all integers $\alpha, \beta\in[1,r]$, such that $\{M_{\alpha}\}_{\alpha=1}^r$ can be simultaneously diagonalized in an orthonormal eigenbasis $\{|i\rangle\}_{i=1}^d$ of the quantum system \cite{Wolf2010,Garcia2017},
\begin{align}\label{Mt}
    %M_{\alpha}=\sum_{m=1}^d c_{\alpha m}|m\rangle\langle m|.
    \left[ {\begin{array}{*{20}{c}}
{{M_1}}\\
 \vdots \\
{{M_r}}
\end{array}} \right] = \left[ {\begin{array}{*{20}{c}}
{{c_{11}}}& \cdots &{{c_{1d}}}\\
 \vdots & \vdots & \vdots \\
{{c_{r1}}}& \cdots &{{c_{rd}}}
\end{array}} \right]\left[ {\begin{array}{*{20}{c}}
{| 1\rangle \langle 1 |}\\
 \vdots \\
{| d \rangle \langle d |}
\end{array}} \right].
\end{align}
This can be simply denoted as $\mathbf{M}=\mathbf{C}\mathbf{P}$, where $\mathbf{M}=[M_1,\cdots,M_r]^T$, $\mathbf{P}=[|1\rangle\langle1|,\cdots,|d\rangle\langle d|]^T$, and $\mathbf{C}$ is a $r\times d$ complex matrix ($r$ and $d$ are generally different). We partition $\mathbf{C}$ according to its columns as $[\mathbf{c}_{1},\cdots, \mathbf{c}_{d}]$, then $\|\mathbf{c}_j\|^2=\mathbf{c}_{j}^{\dagger}\mathbf{c}_{j}=1$ for any integer $j\in[1,d]$ due to $\mathbf{M}^{\dagger}\mathbf{M}=\sum_{\alpha=1}^rM_{\alpha}^{\dagger}M_{\alpha}=\mathbb{I}$, and $\{\mathbf{c}_{j}\}_{j=1}^{d}$ is a set of unit vectors in a $r$-dimensional complex vector space, with $j$ corresponding to the basis state $|j\rangle$. Note that these unit vectors are not necessarily orthogonal to each other \cite{SI}. For a specific GM, the measurement operators are not unique, since we can define a new set of measurement operators by $\mathbf{M}'=\mathbf{T}\mathbf{M}$ with $\mathbf{T}$ being a $r\times r$ unitary matrix, which satisfy $\mathbf{M}'^{\dagger}\mathbf{M}'=\mathbb{I}$ and also characterize the same quantum channel.

The quantum channel is then a diagonal operator acting in the HS space,
\begin{equation}\label{phi}
    \hat{\Phi}=\sum_{i,j=1}^{d}\mathbf{c}_{j}^{\dagger}\mathbf{c}_{i}|ij\rangle\rangle\langle\langle ij|,
\end{equation}
where $\{|ij\rangle\rangle\}_{i,j=1}^{d}$ are the eigenvectors (eigenmatrices in the Hilbert space) of $\hat{\Phi}$ with the corresponding eigenvalues $\{\mathbf{c}_{j}^{\dagger}\mathbf{c}_{i}\}_{i,j=1}^{d}$. Since $|\mathbf{c}_{j}^{\dagger}\mathbf{c}_{i}|\leq1$ (due to the Cauthy-Schwarz inequality) with equality if and only if $\mathbf{c}_{i}=e^{i\varphi}\mathbf{c}_{j}$ for some real $\varphi$, all the eigenvalues of $\hat{\Phi}$ lie within the unit disk of the complex plane. The eigenvectors with eigenvalue 1 are called \textit{fixed points} \cite{Wolf2010,Arias2002}, and those with eigenvalues $e^{i\varphi}$ with $\varphi\neq0$ are \textit{rotating points}. Obviously the fixed points must include $\{|jj\rangle\rangle\}_{j=1}^d$, and the rotating points are $\{|ij\rangle\rangle|\forall i,j\in[1,d], \mathbf{c}_{j}^{\dagger}\mathbf{c}_{i}=e^{i\varphi}\neq1\}$.

As a simple example, consider $\{\mathbf{c}_j\}_{j=1}^d$ as a set of orthonormal vectors, then the channel is $\hat{\Phi}=\sum_{j=1}^{d}|jj\rangle\rangle\langle\langle jj|$, representing a non-selective PM with rank-1 projectors (von Neumann measurements), ${\Phi}(\cdot)=\sum_{j=1}^{d}|j\rangle\langle j|(\cdot)|j\rangle\langle j|$. This channel has only fixed points but no rotating points. As another example, consider $\{\mathbf{c}_j\}_{j=1}^d=\{\widetilde{\mathbf{c}} e^{i\varphi_{j}}\}_{j=1}^d$, then $\hat{\Phi}=\sum_{j=1}^{d}e^{i(\varphi_i-\varphi_j)}|ij\rangle\rangle\langle\langle ij|$ is a unitary channel ${\Phi}(\cdot)=U(\cdot)U^{\dagger}$ with $U=\sum_{j=1}^d e^{i\varphi_j}|j\rangle\langle j|$. For the unitary channel, $|ij\rangle\rangle$ is a fixed point if $i=j$ or $\varphi_i=\varphi_j$, and a rotating point if $\varphi_i\neq\varphi_j$.

For general cases, we divide the index set $A=\{1,\cdots,d\}$ into $s (\leq d)$ disjoint subsets $A_1,\cdots,A_s$, with the corresponding cardinalities (number of elements) being $d_1,\cdots,d_s$, satisfying $\sum_{i=1}^s d_i=d$. Then divide the set of unit vectors $C=\{\mathbf{c}_{j}\}_{j=1}^{d}$ into $s$ disjoint subsets $C_1,\cdots,C_s$ with $C_k=\{\mathbf{c}_{j}|j\in A_k\}$. This division should ensure that the unit vectors in each subset are the same up to some phase factors but are different from any other unit vectors in other subsets, i.e. $C_k=\{\widetilde{\mathbf{c}}_k e^{i\varphi_{j}}|j\in A_k\}$ but $\widetilde{\mathbf{c}}_p \neq\widetilde{\mathbf{c}}_q e^{i\varphi}$ for any $\varphi$ and $p,q\in[1,s]$. This implies that $|ij\rangle\rangle$ with $i,j\in A_k$ is either a fixed point ($\varphi_i=\varphi_j$) or a rotating point ($\varphi_i\neq\varphi_j$).

The division of the index set also partitions the Hilbert space $\mathcal{H}$ of the quantum system into the direct sum of $s$ subspaces, $\mathcal{H}=\mathcal{H}_1\oplus\cdots\oplus\mathcal{H}_s$, where $\mathcal{H}_k={\rm Span}\{|j\rangle|j\in A_k\}$ with rank-$d_k$ projection $P_k=\sum_{j\in A_k}|j\rangle\langle j|$. Thus
the measurement operators in Eq. (\ref{Mt}) can be written in a compact matrix form, $\mathbf{M}=\widetilde{\mathbf{C}}\widetilde{\mathbf{P}}$, where $\widetilde{\mathbf{C}}=[\widetilde{\mathbf{c}}_1,\cdots, \widetilde{\mathbf{c}}_s]$ and $\widetilde{\mathbf{P}}=[\widetilde{P}_1,\cdots, \widetilde{P}_s]^T$ with $\widetilde{P}_k=\sum_{j\in A_k}e^{i\varphi_j}|j\rangle\langle j|$. Note that $\widetilde{P}_k$ is either a projection operator or a unitary operator in $\mathcal{H}_k$, satisfying $\widetilde{P}_k^{\dagger}\widetilde{P}_{k'}=\delta_{kk'}P_k$ and $\sum_{k=1}^s \widetilde{P}_k^{\dagger} \widetilde{P}_k=\mathbb{I}$. Such a compact form of $\mathbf{M}$ allows us to extend the above formulation to infinite-dimensional quantum systems \cite{SI}, if we divide the identity operator into a finite set of orthogonal projections.

\textit{Asymptotics of sequential GMs}. Sequential non-selective GMs correspond to sequential applications of the quantum channel $\hat{\Phi}$ [Fig. \ref{SGM}{\color{blue}(a)}]. Previous works have studied the asymptotic behaviors of sequential general quantum channels  \cite{Albert2019,Burgarth2013,Novotny2018,Blume2010}, mostly trying to find which information from an initial state can be preserved during the process.

For the channel in Eq. (\ref{phi}), as the number of applications $m$ increases, the projections to eigenvectors with eigenvalues lying in the interior of the unit disk ($|\mathbf{c}_{j}^{\dagger}\mathbf{c}_{i}|<1$) gradually vanish, while the projections to eigenvectors with eigenvalues on the unit circle ($|\mathbf{c}_{j}^{\dagger}\mathbf{c}_{i}|=1$) remain unchanged or change by some phase factors. So sequential non-selective GMs tend to preserve the quantum coherence within subspaces $\{\mathcal{H}_k\}_{k=1}^s$ but diminish the coherence between different subspaces.
First assume that the channel has only fixed points, i.e., elements in each $C_k$ are all the same or $\varphi_{j}=0$ for all $j\in [1,d]$, then in the asymptotic limit of large $m$,
\begin{equation}\label{}
    \mathop {\lim }\limits_{m \to \infty} \hat{\Phi}^{m}=\sum_{k=1}^{s}\sum_{i,j\in A_k}|ij\rangle\rangle\langle\langle ij|=\sum_{k=1}^{s}\hat{\mathcal{P}}_k,
\end{equation}
corresponding to $\mathop {\lim }\limits_{m \to \infty} {\Phi}^{m}(\cdot)=\sum_{k=1}^{s}P_k(\cdot)P_k$ [Figs. \ref{SGM}{\color{blue}(b)} and {\color{blue}(c)}], which represents non-selective PMs. Then consider the channel with also rotating points, i.e., there are different phase factors in $C_k=\{\widetilde{\mathbf{c}}_k e^{i\varphi_{j}}|j\in A_k\}$, each application of $\hat{\Phi}$ produces a unitary operation in the Hilbert subspace $\mathcal{H}_k$, i.e., $P_k(\cdot)P_k$ in the former case should be replaced by $\widetilde{P}_k^m(\cdot)(\widetilde{P}_k^{\dagger})^m$. For example, if $C_k=\{\mathbf{c}_i,\mathbf{c}_j\}=\{\widetilde{\mathbf{c}}_k e^{i\varphi_{i}},\widetilde{\mathbf{c}}_ke^{i\varphi_{j}}\}$, then $\widetilde{P}_k=e^{i\varphi_{i}}|i\rangle\langle i|+e^{i\varphi_{j}}|j\rangle\langle j|$. Then the asymptotic limit may not exist but the typicality theory below for finite $m$ still applies in these cases \cite{SI}.

\textit{Typicality of sequential GMs}. Now that sequential non-selective GMs produce projections (or oscillatory unitary operations in the projected subspaces) in the asymptotic limit, we further ask which sequences of sequential selective GMs produce a specific projection. This problem can be perfectly solved by the theory of classical typicality \cite{Wilde2017,Cover2006,Facchi2015,Goldstein2006,Bartsch2009}. Classical typicality mainly cares about the following problem: if a random variable takes $r$ different values with the probability distribution $(p_1,\cdots,p_r)$, generate $m$ independent realizations of this variable and find the statistical distributions of the event sequences with $(m_1/m,\cdots,m_r/m)$, where $m_i$ is the number of the occurrences of the $i$th value. For infinitely large $m$, the event sequences that are overwhelmingly likely to occur are the set of \textit{typical sequences} with $(p_1,\cdots,p_r)$.

A non-selective GM is a quantum instrument, which has $r$ outcomes with an analogous ``probability distribution" $(\hat{\mathcal{M}}_{1}, \cdots, \hat{\mathcal{M}_{r}})$ (note that $\{\hat{\mathcal{M}_{r}}\}_{\alpha=1}^r$ are all diagonal matrices, and their projections to the space of each fixed point defines a probability distribution). For sequential non-selective GMs, we can define sequences of selective GMs [Fig. \ref{SGM}{\color{blue}(a)}]. Below we show that the asymptotic projections are induced by the sets of typical sequences of selective GMs.

Since $\hat{\Phi}=\sum_{\alpha=1}^{r}\hat{\mathcal{M}}_{\alpha}$ and $[\hat{\mathcal{M}}_{\alpha}, \hat{\mathcal{M}}_{\beta}]=0$ for $\alpha, \beta\in[1,r]$, we can expand $\hat{\Phi}^m$ according to the multinomial theorem,
\begin{equation}\label{}
    %\hat{\Phi}^m=\sum_{m_1,\cdots,m_r}\frac{m!}{m_1!m_2!\cdots m_r!}\hat{\mathcal{M}}_{1}^{m_1}\hat{\mathcal{M}}_{2}^{m_2}\cdots\hat{\mathcal{M}}_{r}^{m_r},
    \hat{\Phi}^m=\sum_{\{F\}}\frac{m!}{(mf_1)!\cdots (mf_r!)}\hat{\mathcal{M}}_{1}^{mf_1}\cdots\hat{\mathcal{M}}_{r}^{mf_r},
\end{equation}
where $F=(f_1,\cdots,f_r)$ with $f_i\in[0,1]$ (also a rational number with denominator $m$) satisfying $\sum_{i=1}^rf_i=1$, and the summation is over all distributions $\{F\}$ in a $(r-1)$-dimensional probability space.
For large $m$, $\hat{\Phi}^m$ can be approximated by its projections to the asymptotic subspaces \cite{SI},
\begin{equation}\label{typi}
    \hat{\Phi}^m\approx\sum_{k=1}^s\hat{\mathcal{P}}_k\hat{\Phi}^m\hat{\mathcal{P}}_k\approx \sum_{k=1}^s \sum_{\{F\}} e^{-mS(F\|F_k)}\hat{\mathcal{P}}_k,
\end{equation}
where $F_k=(f_{k1},\cdots,f_{kr})=(|\widetilde{c}_{1k}|^2,\cdots,|\widetilde{c}_{rk}|^2)$ with $\widetilde{c}_{1k}, \cdots, \widetilde{c}_{rk}$ being entries of $\widetilde{\mathbf{c}}_k$ satisfying $\sum_{i=1}^r|\widetilde{c}_{ik}|^2=1$, and $S(F\|F_k)=\sum_{i=1}^r f_i\ln (f_i/f_{ki})$ is the relative entropy between $F$ and $F_k$ (the derivation above uses Stirling's formula $\ln m!\approx m\ln m-m$ for large $m$). $S(F\|F_k)$ takes the minimum when $F=F_k$, so for infinite large $m$, $\{F_k\}_{k=1}^s$ represent $s$ sets of ideal typical sequences of selective GMs leading to the projections $\{\hat{\mathcal{P}}_k\}_{k=1}^s$ correspondingly [Fig. \ref{SGM}{\color{blue}(b)}].

For large but finite $m$, the distributions of selective GM sequences for $\hat{\mathcal{P}}_k$ are concentrated around $F_k$, so $S(F\|F_k)\approx \sum_{i=1}^r(f_i-f_{ki})^2/(2f_{ki})$. Then Eq. (\ref{typi}) represents the summation of $s$ Gaussians around $F_1,\cdots, F_s$, with integration of the $k$th Gaussian over the whole probability space giving rise to $\hat{\mathcal{P}}_k$. For any two Gaussians around $F_j$ and $F_k$, they are well separated if the distance between $F_j$ and $F_k$ is larger than the sum of the respective Gaussian half widths. This requires
$m> 2|\ln \eta|[(\sum_{i=1}^r(f_{ji}-f_{ki})^2/f_{ji})^{-1/2}+(\sum_{i=1}^r(f_{ji}-f_{ki})^2/f_{ki})^{-1/2}]^2$ \cite{SI,Liu2010}, where $\eta$ is the ratio of the minimum hight to the maximum hight within the Gaussian width. If all the Gaussians are well separated, integration of the selective GM sequences within a small neighborhood around $F_k$ can approximate $\hat{\mathcal{P}}_k$ up to arbitrary small error as $m$ increases (see the Supplementary Material \cite{SI} for the error rates with finite $m$).

It may happen that two Gaussians coincide around $F_j=F_k$ but $\widetilde{\mathbf{c}}_j\neq \widetilde{\mathbf{c}}_k$, i.e., only partial elements of $\widetilde{\mathbf{c}}_j$ and $\widetilde{\mathbf{c}}_k$ differ by some phase factors. Since $|\widetilde{\mathbf{c}}_j^{\dagger}\widetilde{\mathbf{c}}_k|<1$, the coinciding Gaussians actually correspond to different projections, and the selective GM sequences around $F_j$ approximately produce $\hat{\mathcal{P}}_j+\hat{\mathcal{P}}_k$. To realize selective projections, we can get a new set of measurement operators by a unitary transformation, thus creating different typical sequences of selective GMs for $\hat{\mathcal{P}}_j$ and $\hat{\mathcal{P}}_k$.

\begin{figure}
\includegraphics[width=3.5in]{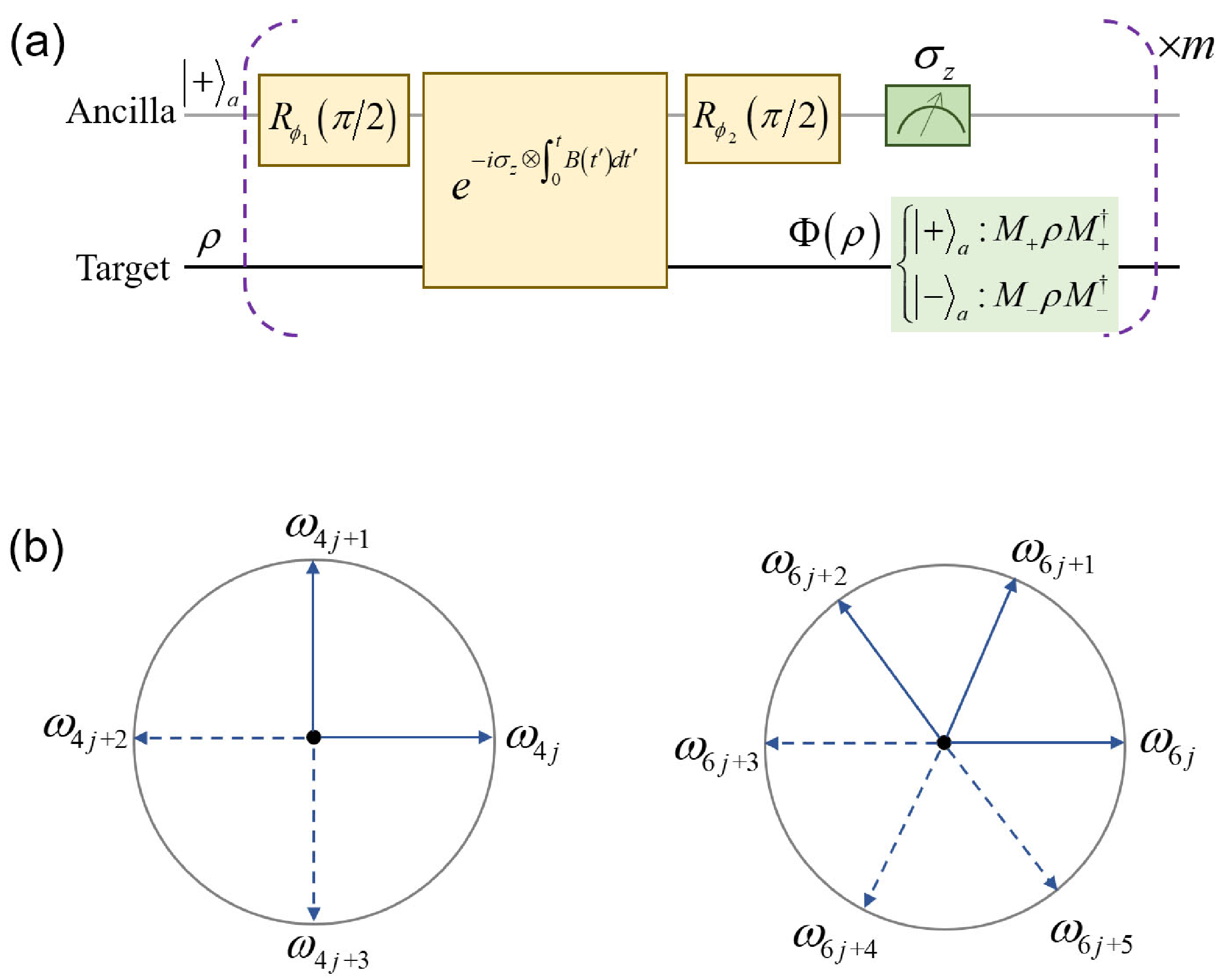}
\caption{(a) Quantum circuit diagram to realize sequential GMs on the target system with PMs of an ancilla qubit. (b) Distributions of eigenvalues of $U_{\pm}(t)=e^{\pm i\chi a^{\dagger}at/2}=\sum_{j=0}^{\infty}\sum_{l=1}^{2N-1} e^{\pm i\omega_{2jN+l}}|2jN+l\rangle\langle 2jN+l|$ in the complex unit circle to detect the $k$ mod $N$ excitation numbers of a bosonic mode with $t=2\pi/(N\chi)$ and $N$=2, 3. }
\label{FigSt}
\end{figure}

\textit{Physical realization}. We present a general physical model to perform PMs on a $d$-level target system with sequential GMs. Without loss of generality, we assume that the GMs are realized by PMs of an ancilla qubit.
The coupling Hamiltonian of the composite system (including the ancilla and target systems) is in the pure-dephasing form \cite{Yang2017}
\begin{equation}\label{}
    H(t)=\sigma_z\otimes B(t),
\end{equation}
where $\sigma_i$ is the Pauli-$i$ operator of the ancilla qubit ($i=x,y,z$), and $B(t)$ is a time-dependent Hermitian operator of the target system (the time-dependence of $B(t)$ is due to being in some interaction picture or external drivings).

The dynamics of the composite system induces a general class of quantum channels on the target system, which can be written in the Stinespring representation as \cite{Stine1955}
\begin{equation}\label{Stine}
    \Phi(\rho)={\rm Tr}_a[U(t)(\rho_a\otimes \rho)U^{\dagger}(t)],
\end{equation}
where $U(t)=\mathcal{T}e^{-i\sigma_z\otimes \int_0^t B(t')dt'}$ with $\mathcal{T}$ being the time-ordering operator, $\rho_a=|\psi\rangle_a\langle\psi|$ is the initial state of the ancilla, $\rho$ denotes the density matrix of the target system, and ${\rm Tr}_a$ denotes the partial trace over the ancilla. With an orthonormal ancilla basis $\{|v_{+}\rangle_a, |v_{-}\rangle_a\}$, we obtain the Kraus representation of the quantum channels, $\Phi(\rho)=\sum_{\alpha\in\{+,-\}}M_{\alpha}\rho M_{\alpha}^{\dagger}$ with $M_{\alpha}=\langle v_{\alpha}|U(t)|\psi\rangle_a$ (note that we add subscripts to the kets only when representing matrix elements or inner products with respect to the ancilla states). With another orthonormal basis $\{T|v_{+}\rangle_a, T|v_{-}\rangle_a\}$ with $T$ being a unitary operator for the ancilla, the measurement operators become $\{M'_{\alpha}\}$ with $M'_{\alpha}=\sum_{\beta\in\{+,-\}} T_{\alpha\beta}M_{\beta}$, while the quantum channels remain unchanged.

We expand $U(t)$ in the ancilla eigenbasis $\{|+\rangle_a, |-\rangle_a\}$ of $\sigma_z$ as $U(t)=|+\rangle_a\langle +|\otimes U_{+}(t)+|-\rangle_a\langle -|\otimes U_{-}(t)$,
%\begin{align}\label{Stine}
%    U(t)=|+\rangle_a\langle +|\otimes U_{+}(t)+|-\rangle_a\langle -|\otimes U_{-}(t),
%\end{align}
where $U_{\pm}(t)=\mathcal{T}e^{\mp i\int_0^t B(t')dt'}$. If $U_{\pm}(t)$ is exactly equal to or well approximated by its first-order Magnus expansion \cite{Ma2016}, i.e., $U_{\pm}(t)=e^{\mp i\int_0^t B(t')dt'}$, then $U_{+}(t)=U_{-}^{\dagger}(t)$ and $[U_{+}(t), U_{-}(t)]=0$, so $U_{+}(t)$ and $U_{-}(t)$ can be simultaneously diagonalized as $U_{\pm}(t)=\sum_{j=1}^d e^{\pm i\omega_j}|j\rangle\langle j|$.
So the measurement operators are $M_{\pm}=\sum_{j=1}^d \left(\langle v_{\pm}|\psi\rangle_a\cos\omega_j + i \langle v_{\pm}|\sigma_z|\psi\rangle_a \sin\omega_j\right)|j\rangle\langle j|$.
As a special case, take $|\psi\rangle_a=R_{\phi_1}(\frac{\pi}{2})|+\rangle_a$ and $|v_{\pm}\rangle_a=R_{\phi_2}(-\frac{\pi}{2})|\pm\rangle_a$ with $R_{\phi}(\theta)=e^{-i(\cos\phi \sigma_x+\sin \phi \sigma_y)\theta/2}$, then
\begin{align}\label{Mpm}
\left[ {\begin{array}{*{20}{c}}
{{{M}_+ }}\\
{{{M}_- }}
\end{array}} \right] = \left[ {\begin{array}{*{20}{c}}
{e^{i\omega_1}-e^{i(\Delta\phi-\omega_1)}}& \cdots &{e^{i\omega_d}-e^{i(\Delta\phi-\omega_d)}}\\
{e^{i\omega_1}+e^{i(\Delta\phi-\omega_1)}}& \cdots &{e^{i\omega_d}+e^{i(\Delta\phi-\omega_d)}}
\end{array}} \right]\frac{\mathbf{P}}{2}
\end{align}
where $\Delta\phi=\phi_1-\phi_2$. Each round of such GMs corresponds to a three-step physical process [Fig. \ref{FigSt}{\color{blue}(a)}]: (1) the ancilla starts from $|+\rangle_a$ and is rotated by $R_{\phi_1}(\frac{\pi}{2})$; (2) let the ancilla and target systems evolve under $H(t)$ for time $t$; (3) finally rotate the ancilla by $R_{\phi_2}(\frac{\pi}{2})$ and make a PM of the ancilla in the basis $\{|+\rangle_a, |-\rangle_a\}$. Similar schemes have been designed to realize single-shot readouts of nuclear spins-1/2 in diamond \cite{Liu2017}, but here we show this scheme can be extended to perform PMs of a generic system.

Since the GMs have only two outcomes, the measurement results are solely determined by the measurement polarization $\Delta f=(m_{-}-m_{+})/m$ \cite{Ma2018}, with $m_{+}/m_{-}$ being the number of outcome $+/-$ in $m$ sequential measurements. For the spectra $\{e^{\pm i\omega_j}\}$ of $U_{\pm}(t)$, calculate $\Delta f_j=\cos(2\omega_j-\Delta\phi)$ for all $j\in[1,d]$. Weak measurement corresponds to the regime $|\Delta f_j|\ll1$. If $\Delta f_j\neq \Delta f_k$ for any $j,k\in[1,d]$ and $j\neq k$, sequential GMs produce von-Neumann measurements of the target system, with the rank-1 projection $P_j=|j\rangle\langle j|$ corresponding to typical selective GM sequences with $\Delta f_j$. If $\Delta f_j=\Delta f_k$, then either (I) $\omega_j+\omega_k=\Delta\phi+n\pi$ or (II) $\omega_j-\omega_k=n\pi$ with $n$ being integers. In case-I, the typical selective GM sequences for $P_j$ and $P_k$ are the same, but selective projections can still be achieved by choosing a different $\Delta\phi'$. In case-II, the typical selective GM sequences with $\Delta f_j$ induce the operation $P_j+(-1)^nP_k$.

\textit{Example: Modular excitation number measurements of bosonic modes}. As an example, we present a protocol to measure the modular excitation numbers of a bosonic mode with an ancilla qubit. The ancilla is dispersively coupled to a bosonic mode with the Hamiltonian $H=-\chi \sigma_z a^{\dagger}a/2$, where $a$ ($a^{\dagger}$) is the annihilation (creation) operator of the bosonic mode and $\chi$ is the dispersive coupling strength. The dispersive coupling arises naturally from the Jaynes-Cumming coupling in cavity quantum electrodynamics (QED) \cite{Raimond2001} and circuit QED \cite{Blais2021} when the detuning between the ancilla and the bosonic mode is much larger than the coupling strength.

We construct the projectors into the sets of bosonic Fock states with modular excitation number $l$ mod $2N$, $P_{2N}^{l}=\sum_{j=0}^{\infty}|2jN+l\rangle\langle 2jN+l|$, with $l\in\{0, 1, \cdots, 2N-1\}$ and $N$ being any positive integer. With the scheme below Eq. (\ref{Mpm}) and the evolution time $t=2\pi/(N\chi)$, $U_{\pm}(t)=e^{\pm i\chi a^{\dagger}at/2}=\sum_{k=0}^{N-1} e^{\pm ik\pi/N}(P_{2N}^{k}-P_{2N}^{k+N})$, i.e. the eigenvalues of $U_{\pm}(t)$ divides the complex unit circle into $2N$ equal pieces [Fig. {\ref{FigSt}{\color{blue}(b)}}]. The measurement operators are $M_{\pm}=\sum_{k=0}^{N-1}(e^{ik\pi/N}\mp e^{i(\Delta\phi-k\pi/N)})(P_{2N}^{k}-P_{2N}^{k+N})$,
and the measurement polarization $\Delta f_k=\cos(2k\pi/N-\Delta\phi)$. We can tune $\Delta\phi$ so that $\Delta f_k$ is maximally distinguishable for different $k\in[0,N-1]$. For $N=1$, $\Delta\phi=0$ is optimal as $\Delta f_0=-\Delta f_1=1$; while for $N\geq2$, we can choose $\Delta\phi=\pi/(2N)$ so that $\Delta f_k=\cos[(2k-1/2)\pi/N]$. Then for a large and even $m$, sequential GMs induce the $k$ mod $N$ excitation number measurement of the bosonic mode. The modular excitation numbers are the error syndromes of rotation-symmetric error correction codes of bosonic modes \cite{Grimsmo2020}, such as cat codes \cite{Leghtas2013,Mirrahimi2014,Li2017,Bergmann2016} and binomial codes \cite{Michael2016}. So this protocol is useful for quantum non-demolition measurements in bosonic quantum information processing \cite{Blais2020,Cai2021,Joshi2021,Ma2021}, especially for tracking the error syndromes of high-order bosonic error correction codes \cite{Sun2014,Ofek2016,Hu2019}.

\textit{Summary}. We have revealed the elegant structures of sequential GMs by studying their asymptotic behaviors and typical sequences. We prove that non-selective PMs can emerge from sequential non-selective GMs when the measurement operators are normal and commuting with each other. Each selective PM comes from a set of typical sequences of selective GMs, which is determined solely by the structures of the measurement operators.
While the GMs here are restricted to have normal and commuting measurement operators, they describe a large class of quantum channels on a quantum system induced by a pure-dephasing coupling between this system and an ancilla system. For future works, it will be interesting to relax this restriction, and study the asymptotics and typicality of sequential GMs with general measurement operators.

\begin{acknowledgments}
W.L.M acknowledges support from Chinese Academy of Sciences (No. E0SEBB11, No. E27RBB11), National Natural Science Foundation of China (No. 12174379, No. E31Q02BG), and Innovation Program for Quantum Science and Technology (No. 2021ZD0302300). R.B.L. was supported by the Hong Kong Research Grants Council - General Research Fund Project 14300119.
\end{acknowledgments}

\textit{Note added}. After completion of this work, we become aware of a related but different work \cite{Linden2022}. In the work of Linden and  Skrzypczyk, they find that with many copies of available GMs in parallel (aided by entangling gates), one can simulate target GMs in the asymptotic limit.

%\end{CJK*}
\end{document}

\title{Supplementary Material for ``Sequential generalized measurements: Asymptotics, typicality and emergent projective measurements''}

\author{Wen-Long Ma}
\email{wenlongma@semi.ac.cn}
\affiliation{State Key Laboratory of Superlattices and Microstructures, Institute of Semiconductors, Chinese Academy of Sciences, Beijing 100083, China}
\affiliation{Center of Materials Science and Opto-Electronic Technology, University of Chinese Academy of Sciences, Beijing 100049, China}
\author{Shu-Shen Li}
\affiliation{State Key Laboratory of Superlattices and Microstructures, Institute of Semiconductors, Chinese Academy of Sciences, Beijing 100083, China}
\affiliation{Center of Materials Science and Opto-Electronic Technology, University of Chinese Academy of Sciences, Beijing 100049, China}
\author{Ren-Bao Liu}
%\email{rbliu@cuhk.edu.hk}
\affiliation{Department of Physics, Centre for Quantum Coherence, and The Hong Kong Institute of Quantum Information Science and Technology, The Chinese University of Hong Kong, Shatin, New Territories, Hong Kong, China}

\date{\today }

%\begin{abstract}
%In this Supplementary Material, we provide an introduction to the basic concepts and details of the derivations in the main text. In Sec. \ref{Sec-HS}, we briefly introduces the Hilbert-Schmidt (HS) space and positive-operator-valued measures (POVMs). Then we provide a systematic description of POVMs with normal and commuting measurement operators in Sec. \ref{Sec-POVM}, and the matrix representation of quantum channels for such POVMs in Sec. \ref{Sec-rep}. In Sec. \ref{Sec-typi}, we give the detailed derivations about determining the typical sequences of POVMs for a specific projective measurement (PM), the conditions to distinguish different typical POVM sequences, and the error rates in approximated PMs with sequential POVMs.
%\end{abstract}

\maketitle

%\tableofcontents
In this Supplementary Material, we provide an introduction to the basic concepts and the details of derivations in the main text. In Sec. \ref{Sec-HS}, we briefly introduce the Hilbert-Schmidt (HS) space and generalized measurements (GMs). Then we provide a systematic description of GMs with normal and commuting measurement operators in Sec. \ref{Sec-POVM}, and the matrix representation of quantum channels for such GMs in Sec. \ref{Sec-rep}. In Sec. \ref{Sec-typi}, we give the detailed derivations about determining the typical sequences of selective GMs for a specific projective measurement (PM), the conditions to distinguish different typical GM sequences, and the error rates in approximating PMs with sequential GMs.

\section{HS space and GMs} \label{Sec-HS}
For a $d$-dimensional quantum system, the space of operators form a linear vector space. This is easily seen if the $d\times d$ complex matrix of an operator $X$ in an orthonormal eigenbasis $\{|i\rangle\}_{i=1}^d$ is reshaped into a $d^2\times 1$ column vector,
\begin{align}\label{}
X = \left[ \begin{array}{l}
\begin{array}{*{20}{c}}
{{x_{11}}}& \cdots &{{x_{1d}}}
\end{array}\\
\begin{array}{*{20}{c}}
 \vdots & \vdots & \vdots \\
{{x_{d1}}}& \cdots &{{x_{dd}}}
\end{array}
\end{array} \right] = \left[ {\begin{array}{*{20}{c}}
{{\mathbf{x}_1}}\\
 \vdots \\
{{\mathbf{x}_d}}
\end{array}} \right], ~~~ \Longleftrightarrow ~~~
\left| {\left. X \right\rangle } \right\rangle  = \left[ {\begin{array}{*{20}{c}}
{\mathbf{x}_1^T}\\
 \vdots \\
{\mathbf{x}_d^T}
\end{array}} \right],
\end{align}
where $\mathbf{x}_i$ is the $i$th row of $X$ with $i\in[1,d]$, and $\mathbf{x}_i^T$ is the transpose of $\mathbf{x}_i$. With Dirac notations, the matrix reshaping can be simply represented by $X=\sum_{i,j=1}^d x_{ij}|i\rangle\langle j|\leftrightarrow |X\rangle\rangle=\sum_{i,j=1}^d x_{ij}|ij\rangle\rangle$. Then the ordinary scalar product between $|X\rangle\rangle$ and $|Y\rangle\rangle$ defines an inner product between $X$ and $Y$,
\begin{align}\label{}
    \langle\langle Y|X\rangle\rangle=\sum_{i=1}^d \mathbf{y}_i^* \mathbf{x}_i^T=\sum_{i,j=1}^d y_{ij}^{*}x_{ij}={\rm Tr}(Y^{\dagger}X),
\end{align}
which is the so-called Hilbert-Schmidt (HS) inner product. The HS space is the space of operators equipped with the HS inner product.

The density matrices of the quantum system, as the class of positive operators with trace one, are also vectors in the HS space. In the HS space, the trace one constraint of a density matrix $\rho$ is equivalent to $\langle\langle \mathbb{I}|\rho\rangle\rangle={\rm Tr}(\rho)=1$, with $\mathbb{I}$ being the identity operator. Left and right multiplications of $\rho$ by operators $X$ and $Y$ corresponds to multiplying $|\rho\rangle\rangle$ with a $d^2\times d^2$ matrix,
\begin{align}\label{}
    X\rho Y=\sum_{i,j=1}^d x_{ik}y_{lj}\rho_{kl}|i\rangle\langle j|, ~~~ \Longleftrightarrow ~~~ X\otimes Y^{T}|\rho\rangle\rangle.
\end{align}
So the operation $X(\cdot)Y$ as a superoperator acting in the Hilbert space is equivalent to a linear operator $X\otimes Y^{T}$ acting in the HS space.

For the quantum channel of a non-selective GM with the measurement operators $\{M_{\alpha}\}_{\alpha=1}^r$, the transformation from its Kraus representation in the Hilbert space to the matrix representation in the HS space is given by
\begin{align}\label{}
    \Phi(\rho)=\sum_{\alpha=1}^{r}\mathcal{M}_{\alpha}(\rho)=\sum_{\alpha=1}^{r}M_{\alpha}\rho M_{\alpha}^{\dagger}, ~~~\Longleftrightarrow ~~~%\\%\Leftrightarrow
    %&\Updownarrow  \nonumber \\
    \hat{\Phi}|\rho\rangle\rangle=\sum_{\alpha=1}^{r}\hat{\mathcal{M}}_{\alpha}|\rho\rangle\rangle=\sum_{\alpha=1}^{r}M_{\alpha}\otimes M_{\alpha}^{*}|\rho\rangle\rangle.
\end{align}
where $\mathcal{M}_{\alpha}=M_{\alpha}(\cdot)M_{\alpha}^{\dagger}$ is a superoperator acting in the Hilbert space representing a selective GM with the $\alpha$th outcome, $\sum_{\alpha=1}^r M_{\alpha}^{\dagger}M_{\alpha}=\mathbb{I}$ ensures the trace-preserving property, $\hat{\mathcal{M}}_{\alpha}=M_{\alpha}\otimes M_{\alpha}^{*}$ is an operator acting in the HS space corresponding to $\mathcal{M}_{\alpha}$, and $M_{\alpha}^{*}$ is the complex conjugate of $M_{\alpha}$. For a selective GM with the $\alpha$th outcome, the density matrix undergoes the following evolution,
\begin{align}\label{}
    \rho_{\alpha}=\frac{\mathcal{M}_{\alpha}(\rho)}{p_{\alpha}}=\frac{M_{\alpha}\rho M_{\alpha}^{\dagger}}{p_{\alpha}}, ~~~\Longleftrightarrow ~~~
     |\rho_{\alpha}\rangle\rangle=\frac{\hat{\mathcal{M}}_{\alpha}|\rho\rangle\rangle}{p_{\alpha}}=\frac{M_{\alpha}\otimes M_{\alpha}^{*}|\rho\rangle\rangle}{p_{\alpha}},
\end{align}
where $p_{\alpha}={\rm Tr}(M_{\alpha}\rho M_{\alpha}^{\dagger})=\langle\langle \mathbb{I}|\hat{\mathcal{M}}_{\alpha}|\rho\rangle\rangle$ is the probability to get the $\alpha$th outcome, satisfying $\sum_{\alpha=1}^r p_{\alpha}=1$.

\section{GMs with normal and commuting measurement operators}\label{Sec-POVM}
In this section, we provide a systematic description of GMs with normal and commuting measurement operators $\{M_{\alpha}\}_{\alpha=1}^r$. Since $[M_{\alpha}, M_{\alpha}^{\dagger}]=[M_{\alpha}, M_{\beta}]=0$ for all integers $\alpha, \beta\in[1,r]$, $\{M_{\alpha}\}_{\alpha=1}^r$ can be simultaneously diagonalized in an orthonormal eigenbasis $\{|i\rangle\}_{i=1}^d$ of the quantum system \cite{Garcia2017},
\begin{align}\label{Mt}
    %M_{\alpha}=\sum_{m=1}^d c_{\alpha m}|m\rangle\langle m|.
    \left[ {\begin{array}{*{20}{c}}
{{M_1}}\\
 \vdots \\
{{M_r}}
\end{array}} \right] = \left[ {\begin{array}{*{20}{c}}
{{c_{11}}}& \cdots &{{c_{1d}}}\\
 \vdots & \vdots & \vdots \\
{{c_{r1}}}& \cdots &{{c_{rd}}}
\end{array}} \right]\left[ {\begin{array}{*{20}{c}}
{| 1\rangle \langle 1 |}\\
 \vdots \\
{| d \rangle \langle d |}
\end{array}} \right],
\end{align}
which can be written in a matrix form as $\mathbf{M}=\mathbf{C}\mathbf{P}$, with the definitions below
\begin{align}\label{}
\mathbf{M} = \left[{\begin{array}{*{20}{c}}
{{M_1}}\\
 \vdots \\
{{M_r}}
\end{array}}\right], ~~
\mathbf{C}= [{{\bf{c}}_1},\;\cdots \;,{{\bf{c}}_d}]=\left[ {\begin{array}{*{20}{c}}
{{c_{11}}}& \cdots &{{c_{1d}}}\\
 \vdots & \vdots & \vdots \\
{{c_{r1}}}& \cdots &{{c_{rd}}}
\end{array}} \right],~~
\mathbf{P}=\left[ {\begin{array}{*{20}{c}}
{| 1 \rangle \langle 1|}\\
 \vdots \\
{| d \rangle \langle d|}
\end{array}} \right]
\end{align}
where $\mathbf{M}$ is a $r\times 1$ column vector of operators, $\mathbf{C}$ is a $r\times d$ complex matrix with $\mathbf{c}_i$ being its $i$th column, and $\mathbf{P}$ is a $d\times 1$ column vector of operators. Further define $\mathbf{M}^{\dagger}=[M_1^{\dagger},\cdots,M_r^{\dagger}]$ and $\mathbf{P}^{\dagger}=\mathbf{P}^{T}=[|1\rangle\langle 1|,\cdots,|d\rangle\langle d|]$, then $\mathbf{M}^{\dagger}\mathbf{M}=\sum_{i=1}^r M_i^{\dagger}M_i=\mathbf{P}^{\dagger}\mathbf{P}=\sum\limits_{i= 1}^d| i\rangle \langle i|=\mathbb{I}$. This condition restricts the form of $\mathbf{C}$, as can be seen by
\begin{align}\label{}
    {{\bf{M}}^\dag }{\bf{M}} =\mathbf{P}^{\dagger}\mathbf{C}^{\dagger}\mathbf{C}\mathbf{P}= \left[ {| 1\rangle \langle1|, \cdots, | d \rangle \langle d|} \right]\left[ {\begin{array}{*{20}{c}}
{{\bf{c}}_1^\dag {{\bf{c}}_1}}& \cdots &{{\bf{c}}_1^\dag {{\bf{c}}_d}}\\
 \vdots & \vdots & \vdots \\
{{\bf{c}}_d^\dag {{\bf{c}}_1}}& \cdots &{{\bf{c}}_d^\dag {{\bf{c}}_d}}
\end{array}} \right]\left[ {\begin{array}{*{20}{c}}
{| 1 \rangle \langle 1|}\\
 \vdots \\
{| d\rangle \langle d|}
\end{array}} \right]
= \sum\limits_{i,j = 1}^d {{\bf{c}}_i^\dag {{\bf{c}}_j}} |i\rangle\langle i|j\rangle \langle j| = \sum\limits_{i= 1}^d {{\bf{c}}_i^\dag {{\bf{c}}_i}} | i\rangle \langle i|,
\end{align}
which clearly shows ${\bf{c}}_i^\dag {{\bf{c}}_i}=\sum_{j=1}^d |c_{ij}|^2=1$ for any $i\in[1,d]$, i.e., all the columns $\{\mathbf{c}_{j}\}_{j=1}^{d}$ of $\mathbf{C}$ are unit vectors in a $r$-dimensional complex vector space. But these unit vectors are not necessarily orthogonal to each other. The reason is that entries of $\mathbf{P}$ are not real or complex numbers but rank-1 projectors $\{|i\rangle\langle i|\}_{i=1}^d$, satisfying $|i\rangle\langle i|j\rangle\langle j|=\delta_{ij}|i\rangle\langle i|$.

Now we take a closer look at the structures of matrix $\mathbf{C}$. Define the set of its column vectors as $C=\{\mathbf{c}_{j}\}_{j=1}^{d}$ with a index set $A=\{1,\cdots,d\}$. Then divide $C$ into $s$ disjoint subsets $C_1,\cdots,C_s$ with the corresponding index subsets $A_1,\cdots,A_s$, where $C_k=\{\mathbf{c}_{j}|j\in A_k\}$ for any integer $k\in[1,s]$. The cardinality of $C_k$ and $A_k$ is $d_k$, satisfying $\sum_{i=1}^s d_i=d$ and $d_i\geq1$. This division should ensure that the unit vectors in each subset are the same up to some phase factors but are different from any other unit vectors in other subsets, i.e. $C_k=\{\widetilde{\mathbf{c}}_k e^{i\varphi_{j}}|j\in A_k\}$ but $\widetilde{\mathbf{c}}_p \neq\widetilde{\mathbf{c}}_q e^{i\varphi}$ for any real $\varphi$ and $p,q\in[1,s]$. This means that we can always simultaneously reorder the columns of $C$ and the entries of $\mathbf{P}$, and then relabel the eigenbasis $\{|i\rangle\}_{i=1}^d$, so that $\mathbf{\mathbf{C}}$ is in the following canonical form,
\begin{align}\label{}
\mathbf{C} = [\overbrace {{{{\bf{\tilde c}}}_1}{e^{i{\varphi _1}}}, \cdots ,{{{\bf{\tilde c}}}_1}{e^{i{\varphi _{{d_1}}}}}}^{{d_1}},\;\overbrace {{{{\bf{\tilde c}}}_2}{e^{i{\varphi _{{d_1} + 1}}}}, \cdots ,{{{\bf{\tilde c}}}_2}{e^{i{\varphi _{{d_1} + {d_2}}}}}}^{{d_2}},\; \cdots  \cdots ,\;\;\overbrace {{{{\bf{\tilde c}}}_s}{e^{i{\varphi _{d - {d_s} + 1}}}}, \cdots ,{{{\bf{\tilde c}}}_s}{e^{i{\varphi _{d - 1}}}}}^{{d_s}}],
\end{align}
and $\mathbf{P}$ remains unchanged. Then Eq. (\ref{Mt}) can be written in a more compact matrix form, $\mathbf{M}=\widetilde{\mathbf{C}}\widetilde{\mathbf{P}}$, where $\widetilde{\mathbf{C}}=[\widetilde{\mathbf{c}}_1,\cdots, \widetilde{\mathbf{c}}_s]$ and $\widetilde{\mathbf{P}}=[\widetilde{P}_1,\cdots, \widetilde{P}_s]^T$ with $\widetilde{P}_k=\sum_{j\in A_k}e^{i\varphi_j}|j\rangle\langle j|$. More explicitly,
\begin{align}\label{Mtc}
    %M_{\alpha}=\sum_{m=1}^d c_{\alpha m}|m\rangle\langle m|.
    \left[ {\begin{array}{*{20}{c}}
{{M_1}}\\
 \vdots \\
{{M_r}}
\end{array}} \right] = \left[ {\begin{array}{*{20}{c}}
{{\widetilde{c}_{11}}}& \cdots &{{\widetilde{c}_{1s}}}\\
 \vdots & \vdots & \vdots \\
{{\widetilde{c}_{r1}}}& \cdots &{{\widetilde{c}_{rs}}}
\end{array}} \right]\left[ {\begin{array}{*{20}{c}}
{\widetilde{P}_1}\\
 \vdots \\
{\widetilde{P}_s}
\end{array}} \right].
\end{align}
While Eq. (\ref{Mt}) mainly concerns about finite-dimensional quantum systems, Eq. (\ref{Mtc}) can describe both finite- and infinite-dimensional systems. The key point is to first partition the identity operator $\mathbb{I}$ of a generic system into a set of orthogonal projections $\{P_k\}_{k=1}^s$, which corresponds to partitioning the Hilbert space $\mathcal{H}$ of the system into the direct sum of $s$ subspaces, $\mathcal{H}=\mathcal{H}_1\oplus\cdots\oplus\mathcal{H}_s$. Then $\widetilde{P}_k$ is a unitary operator in subspace $\mathcal{H}_k$ (with projection $P_k$ as a special case), satisfying $\widetilde{P}_k^{\dagger}\widetilde{P}_{k'}=\delta_{kk'}P_k$ and $\widetilde{\mathbf{P}}^{\dagger}\widetilde{\mathbf{P}}=\sum_{k=1}^s \widetilde{P}_k^{\dagger} \widetilde{P}_k=\mathbb{I}$. Obviously, projective measurements and unitary evolutions are both special cases of Eq. (\ref{Mtc}).

Moreover, for both Eq. (\ref{Mt}) and Eq. (\ref{Mtc}), we can define a new set of measurement operators by $\mathbf{M}'=\mathbf{T}\mathbf{M}$ with $\mathbf{T}=[T_{\alpha\beta}]$ being a $r\times r$ unitary matrix, which satisfies $\mathbf{M}'^{\dagger}\mathbf{M}'=\mathbf{M}\mathbf{T}^{\dagger}\mathbf{T}\mathbf{M}=\mathbb{I}$. $\mathbf{M}'$ and $\mathbf{M}$ also characterize the same CPTP map, since
\begin{align}\label{}
    \sum_{\alpha=1}^{r} M_{\alpha}' (\cdot) M_{\alpha}'^{\dagger}=\sum_{\alpha,\beta,\gamma=1}^{r}T_{\alpha\gamma}^{*}T_{\alpha\beta}M_{\beta} (\cdot) M_{\gamma}^{\dagger}=\sum_{\beta,\gamma=1}^{r}\delta_{\gamma\beta}M_{\beta} (\cdot) M_{\gamma}^{\dagger}=\sum_{\beta=1}^{r}M_{\beta} (\cdot) M_{\beta}^{\dagger}.
\end{align}
where we have used $\sum_{\alpha=1}^{r}T_{\alpha\gamma}^{*}T_{\alpha\beta}=\delta_{\gamma\beta}$ since $\mathbf{T}$ is a unitary matrix.

\section{Representations of quantum channels for GMs}\label{Sec-rep}

For the measurement operators in Eq. (\ref{Mt}), the matrix representation of the channel acting in the HS space is
\begin{align}\label{}
    \hat{\Phi}=\sum_{\alpha=1}^{r}\sum_{i,j=1}^{d}(c_{\alpha i}|i\rangle\langle i|)\otimes (c_{\alpha j}^{*}|j\rangle\langle j|)=\sum_{i,j=1}^{d}\left(\sum_{\alpha=1}^{r}c_{\alpha j}^{*}c_{\alpha i}\right)|ij\rangle\rangle\langle\langle ij|=\sum_{i,j=1}^{d}\mathbf{c}_{j}^{\dagger}\mathbf{c}_{i}|ij\rangle\rangle\langle\langle ij|.
\end{align}
while for the more general case in Eq. (\ref{Mtc}), we can similarly obtain
\begin{align}\label{}
    \hat{\Phi}=\sum_{\alpha=1}^{r}\sum_{k,l=1}^{s}(\widetilde{c}_{\alpha k}\widetilde{P}_k)\otimes (\widetilde{c}_{\alpha l}^{*}\widetilde{P}_l^{*})=\sum_{k,l=1}^{s}\widetilde{\mathbf{c}}_{l}^{\dagger}\widetilde{\mathbf{c}}_{k}\left(\widetilde{P}_k\otimes\widetilde{P}_l^{*}\right)
    =\sum_{k,l=1}^{s}\widetilde{\mathbf{c}}_{l}^{\dagger}\widetilde{\mathbf{c}}_{k}\left(\sum_{i\in A_k}\sum_{j\in A_l}e^{i(\varphi_i-\varphi_j)}|ij\rangle\rangle\langle\langle ij|\right).
\end{align}
where $\{\widetilde{P}_k\otimes\widetilde{P}_l^{*}\}_{k,l=1}^s$ is a set of $s^2$ diagonal matrices in HS space that has orthogonal supports, i.e., $(\widetilde{P}_k\otimes\widetilde{P}_l^{*})(\widetilde{P}_{k'}\otimes\widetilde{P}_{l'}^{*})=\delta_{kk'}\delta_{ll'}(\widetilde{P}_k\otimes\widetilde{P}_l^{*})^2$. From the Cauthy-Schwarz inequality, $\widetilde{\mathbf{c}}_{l}^{\dagger}\widetilde{\mathbf{c}}_{k}<1$ if $k\neq l$, since $\widetilde{\mathbf{c}}_{k}\neq \widetilde{\mathbf{c}}_{l}e^{i\varphi}$ for any real $\varphi$. So with many applications of the channel,
\begin{align}\label{}
    \hat{\Phi}^{m}=\sum_{k,l=1}^{s}(\widetilde{\mathbf{c}}_{l}^{\dagger}\widetilde{\mathbf{c}}_{k})^m\left(\widetilde{P}_k\otimes\widetilde{P}_l^{*}\right)^m\approx \sum_{k=1}^{s}\left(\widetilde{P}_k\otimes\widetilde{P}_k^{*}\right)^m=\sum_{k=1}^{s}\sum_{i,j\in A_k}e^{im(\varphi_i-\varphi_j)}|ij\rangle\rangle\langle\langle ij|.
\end{align}
If $\varphi_j=0$ for any $j\in[1,d]$, i.e., $\widetilde{P}_k=P_k$, then
\begin{align}\label{}
    \hat{\Phi}^{m}\approx\sum_{k=1}^{s}P_k\otimes P_k=\sum_{k=1}^{s}\hat{\mathcal{P}}_k=\sum_{k=1}^{s}\sum_{i,j\in A_k}|ij\rangle\rangle\langle\langle ij|.
\end{align}

\section{Typicality of sequential GMs}\label{Sec-typi}
Since $[M_{\alpha}, M_{\beta}]=0$ for $\alpha, \beta\in[1,r]$, we can easily prove that $[\hat{\mathcal{M}}_{\alpha}, \hat{\mathcal{M}}_{\beta}]=0$. So $\hat{\Phi}^m$ can be expanded according to the multinomial theorem,
\begin{align}\label{}
    \hat{\Phi}^m=\left(\sum_{\alpha=1}^{r}\hat{\mathcal{M}}_{\alpha}\right)^m=\sum_{\alpha_1,\cdots,\alpha_m=1}^r \hat{\mathcal{M}}_{\alpha_1}\cdots \hat{\mathcal{M}}_{\alpha_m}=\sum_{\scriptstyle ~m_1,\cdots,m_r\geq0\hfill\atop
\scriptstyle m_1+\cdots+m_r=m\hfill}\frac{m!}{m_1!m_2!\cdots m_r!}\hat{\mathcal{M}}_{1}^{m_1}\hat{\mathcal{M}}_{2}^{m_2}\cdots\hat{\mathcal{M}}_{r}^{m_r}.
\end{align}
We define a distribution $F=(f_1,\cdots,f_r)=(m_1/m,\cdots,m_r/m)$ to represent the frequencies of each superoperator in $\{\hat{\mathcal{M}}_{\alpha}\}_{\alpha=1}^r$ to appear in the POVM sequence $\hat{\mathcal{M}}_{\alpha_1}\cdots \hat{\mathcal{M}}_{\alpha_m}$, where $\sum_{i=1}^r f_i=1$. Then $\hat{\Phi}^m$ can be rewritten as
\begin{align}\label{phiF}
    \hat{\Phi}^m=\sum_{\{F\}}\frac{m!}{(mf_1)!\cdots (mf_r!)}\hat{\mathcal{M}}_{1}^{mf_1}\cdots\hat{\mathcal{M}}_{r}^{mf_r},
\end{align}
where the summation is over all distributions $\{F\}$ in a $(r-1)$-dimensional probability space.
Substituting $\hat{\mathcal{M}}_{\alpha}=\sum_{k,l=1}^{s}\widetilde{c}_{\alpha k}\widetilde{c}_{\alpha l}^{*}(\widetilde{P}_k\otimes\widetilde{P}_l^{*})$ into Eq. (\ref{phiF}) gives
\begin{align}\label{phiF2}
    \hat{\Phi}^m=&\sum_{k,l=1}^{s}\sum_{\{F\}}\frac{m!}{(mf_1)!\cdots (mf_r!)}(\widetilde{c}_{1 k}\widetilde{c}_{1 l}^{*})^{mf_1}\cdots
    (\widetilde{c}_{r k}\widetilde{c}_{r l}^{*})^{mf_r}(\widetilde{P}_k\otimes\widetilde{P}_l^{*})^m  \nonumber \\
    \approx&\sum_{k=1}^{s}\sum_{\{F\}}\frac{m!}{(mf_1)!\cdots (mf_r!)}|\widetilde{c}_{1 k}|^{2mf_1}\cdots
    |\widetilde{c}_{r k}|^{2mf_r}(\widetilde{P}_k\otimes\widetilde{P}_k^{*})^m,
\end{align}
where we have used $(\widetilde{P}_k\otimes\widetilde{P}_l^{*})(\widetilde{P}_{k'}\otimes\widetilde{P}_{l'}^{*})=\delta_{kk'}\delta_{ll'}(\widetilde{P}_k\otimes\widetilde{P}_l^{*})^2$.
To further simplify Eq. (\ref{phiF2}), we can use Stirling's formula $\ln m!\approx m\ln m-m$ for large $m$ to obtain
\begin{align}\label{}
    \ln \left( \frac{m!}{(mf_1)!\cdots (mf_r!)}|\widetilde{c}_{1 k}|^{2mf_1} \cdots
    |\widetilde{c}_{r k}|^{2mf_r}\right)\approx -m\sum_{i=1}^r f_i \ln\frac{f_i}{|\widetilde{c}_{i k}|^{2}}=-mS(F\|F_k),
\end{align}
where we define $F_k=(f_{k1},\cdots,f_{kr})=(|\widetilde{c}_{1k}|^2,\cdots,|\widetilde{c}_{rk}|^2)$ with $\widetilde{c}_{1k}, \cdots, \widetilde{c}_{rk}$ being entries of $\widetilde{\mathbf{c}}_k$ satisfying $\sum_{i=1}^r|\widetilde{c}_{ik}|^2=1$, and $S(F\|F_k)=\sum_{i=1}^r f_i\ln (f_i/f_{ki})$ is the relative entropy between $F$ and $F_k$. Then Eq. (\ref{phiF2}) is reduced to
\begin{align}\label{phiF3}
    \hat{\Phi}^m\approx \sum_{k=1}^s \sum_{\{F\}} e^{-mS(F\|F_k)}(\widetilde{P}_k\otimes\widetilde{P}_k^{*})^m.
\end{align}
Moreover, for relatively large $m$, the distribution $e^{-mS(F\|F_k)}$ is concentrated within a small neighborhood around $F_k$, so $S(F\|F_k)\approx \sum_{i=1}^r(f_i-f_{ki})^2/(2f_{ki})$, and Eq. (\ref{phiF2}) can be further approximated as
\begin{align}\label{phiF4}
    \hat{\Phi}^m\approx \sum_{k=1}^s \sum_{\{F\}} e^{-\frac{m}{2}\sum_{i=1}^r\frac{(f_i-f_{ki})^2}{f_{ki}}}(\widetilde{P}_k\otimes\widetilde{P}_k^{*})^m.
\end{align}

For the special case $\widetilde{P}_k=P_k$, Eq. (\ref{phiF3}) and Eq. (\ref{phiF4}) become
\begin{align}\label{phiF4}
    \hat{\Phi}^m\approx \sum_{k=1}^s \sum_{\{F\}} e^{-mS(F\|F_k)}\hat{\mathcal{P}}_k\approx\sum_{k=1}^s \sum_{\{F\}} e^{-\frac{m}{2}\sum_{i=1}^r\frac{(f_i-f_{ki})^2}{f_{ki}}}\hat{\mathcal{P}}_k,
\end{align}
which represents summations of $s$ Gaussians around $F_1,\cdots, F_s$, with integration of the $k$th Gaussian over the whole probability space giving rise to $\hat{\mathcal{P}}_k$.

For any two Gaussians around $F_j$ and $F_k$, they are well separated if the distance between $F_j$ and $F_k$ is larger than the sum of the respective Gaussian half widths. In the $(r-1)$-dimensional probability space, the straight line connecting $F_j$ and $F_k$ is
\begin{align}\label{}
    F_{jk}(t)=(1-t)F_j+tF_k=((1-t)f_{j1}+tf_{k1},\cdots, (1-t)f_{jr}+tf_{kr}),
\end{align}
where $t$ is a real number within $[0,1]$. Define $\eta$ as the ratio of the minimum hight to the maximum hight within the Gaussian width, then the half widths $\Delta t_j$ and $\Delta t_k$ of the two Gaussians along the line $F_{jk}(t)$ can be derived as
\begin{align}\label{}
    &e^{-\frac{m}{2}\sum_{i=1}^r\frac{(\Delta t_j)^2(f_i-f_{ji})^2}{f_{ji}}}=\eta, ~~~\Longrightarrow~~~ \Delta t_j=\sqrt{\frac{2|\ln\eta|}{m}}\left( \sum_{i=1}^r\frac{(f_{ji}-f_{ki})^2}{f_{ji}} \right)^{-\frac{1}{2}}, \\
    & e^{-\frac{m}{2}\sum_{i=1}^r\frac{(\Delta t_k)^2(f_i-f_{ji})^2}{f_{ki}}}=\eta, ~~~\Longrightarrow~~~ \Delta t_k=\sqrt{\frac{2|\ln\eta|}{m}}\left( \sum_{i=1}^r\frac{(f_{ji}-f_{ki})^2}{f_{ki}} \right)^{-\frac{1}{2}}, \\
\end{align}
so the two Gaussians around $F_j$ and $F_k$ are well separated if
\begin{align}\label{}
    \Delta t_j+\Delta t_k<1, ~~~\Longrightarrow~~~ m>2|\ln\eta|\left[ \left( \sum_{i=1}^r\frac{(f_{ji}-f_{ki})^2}{f_{ji}} \right)^{-\frac{1}{2}}+ \left( \sum_{i=1}^r\frac{(f_{ji}-f_{ki})^2}{f_{ki}} \right)^{-\frac{1}{2}} \right]^2.
\end{align}
For all the Gaussians to be well separated, the lower bound for the number of measurements is
\begin{align}\label{}
    m>2|\ln\eta|\mathop {\max }\limits_{\scriptstyle ~~j \ne k \hfill \atop
\scriptstyle j,k \in [1,s]\hfill} \left[ \left( \sum_{i=1}^r\frac{(f_{ji}-f_{ki})^2}{f_{ji}} \right)^{-\frac{1}{2}}+ \left( \sum_{i=1}^r\frac{(f_{ji}-f_{ki})^2}{f_{ki}} \right)^{-\frac{1}{2}} \right]^2.
\end{align}

For the $j$th Gaussian, we define a closed neighborhood $\mathcal{F}_j^{\delta}$ around $F_j$ in the probability space, such that summation of all the selective GM sequences within $\mathcal{F}_j^{\delta}$ well approximates $\hat{\mathcal{P}}_j$. Explicitly, summation of all the selective GM sequences within $\mathcal{F}_j^{\delta}$ gives
\begin{align}\label{}
    \hat{\mathcal{P}}_j^{\delta}\approx\sum_{k=1}^s\sum_{F\in\mathcal{F}_j^{\delta}} e^{-mS(F\|F_k)}\hat{\mathcal{P}}_k=\sum_{k=1}^s w_{jk} \hat{\mathcal{P}}_k.
\end{align}
where $w_{jk}=\sum_{F\in\mathcal{F}_j^{\delta}} e^{-mS(F\|F_k)}$, satisfying $\sum_{j=1}^sw_{jk}\leq1$ and $w_{jk}\geq 0$. Define $F_j^{*}$ as a point on the boundary of $\mathcal{F}_j^{\delta}$ where $F$ takes the minimum on the boundary, then from classical typicality theory \cite{Cover2006,Mardia2020}, we have
\begin{align}\label{}
    1-w_{jj}<\frac{(m+r-1)!}{m!(r-1)!}e^{-mS(F\|F_j^{*})}.
\end{align}
As $\hat{\mathcal{P}}_j^{\delta}$ and $\hat{\mathcal{P}}_j$ are both diagonal operators in the HS space, we can use the trace distance to estimate an upper bound of the error rate in approximating $\hat{\mathcal{P}}_j$ with $\hat{\mathcal{P}}_j^{\delta}$,
\begin{align}\label{}
    ||\hat{\mathcal{P}}_j^{\delta}-\hat{\mathcal{P}}_j||_1\approx 1-w_{jj}+\sum_{k\neq j}w_{jk}\leq \sum_{k=1}^s(1-w_{kk})<\sum_{k=1}^s\frac{(m+r-1)!}{m!(r-1)!}e^{-mS(F\|F_k^{*})},
\end{align}
which can be made arbitrarily small for a large enough $m$.

%\begin{align}\label{}

%\end{align}